\documentclass[preprint,showpacs,preprintnumbers,amsmath,amssymb]{revtex4}
\pdfoutput=1
\usepackage{graphicx}% Include figure files
\usepackage{dcolumn}% Align table columns on decimal point
\usepackage{bm}% bold math

\begin{document}

%\preprint{MIT/MKI-GETANUMBER}

\title{First Dark Matter Search Results from a Surface Run of the 10-L DMTPC Directional Dark Matter Detector}

\author{(DMTPC Collaboration) S. Ahlen$^{4}$, J. B. R. Battat$^{6}$, T. Caldwell$^{6}$, C. Deaconu$^{6}$, D. Dujmic$^{1, 6}$, W. Fedus$^{6}$, P. Fisher$^{1, 2, 6}$, F. Golub$^{5}$, S. Henderson$^{6}$, A. Inglis$^{4}$, A. Kaboth$^{6}$, G. Kohse$^{3}$, R. Lanza$^{3}$, A. Lee$^{6}$, J. Lopez$^{6}$, J. Monroe$^{6}$, T. Sahin$^{6}$, G. Sciolla$^{6}$, N. Skvorodnev$^{5}$, H. Tomita$^{4}$, H. Wellenstein$^{5}$, I. Wolfe$^{6}$, R. Yamamoto$^{6}$, H. Yegoryan$^{6}$}
\affiliation{
$^1$Laboratory for Nuclear Science, Massachusetts Institute of Technology; Cambridge, MA 02139 \\
$^2$MIT Kavli Institute for Astrophysics and Space Research, Massachusetts Institute of Technology; Cambridge, MA 02139 \\
$^3$Nuclear Science and Engineering Department, Massachusetts Institute of Technology; Cambridge, MA 02139 \\
$^4$Physics Department, Boston University; Boston, MA 02215 \\
$^5$Physics Department, Brandeis University; Waltham, MA 02453\\
$^6$Physics Department, Massachusetts Institute of Technology; Cambridge, MA 02139
}
\email{jmonroe@mit.edu}
\date{\today}% It is always \today, today,

\begin{abstract} The Dark Matter Time Projection Chamber (DMTPC) is a
low pressure (75 Torr CF$_{4}$) 10 liter detector capable of measuring
the vector direction of nuclear recoils with the goal of directional
dark matter detection.  In this paper we present the first dark matter
limit from DMTPC from a surface run at MIT.  In an analysis window of
80-200 keV recoil energy, based on a 35.7 g-day exposure, we set a
90\% C.L. upper limit on the spin-dependent WIMP-proton cross section
of $2.0\times10^{-33}$ cm$^{2}$ for 115 GeV/c$^2$ dark matter particle
mass.  \end{abstract}

\maketitle

\section{Introduction}
\vspace{-0.3cm}
 Despite strong astrophysical evidence that dark matter comprises
 approximately 23$\%$ of our
 universe~\cite{Spergel:2006hy}, the nature of this dark matter
 remains largely unknown.  Weakly interacting massive particles
 (WIMPs) are a favored dark matter candidate
~\cite{Ellis:2000jd}.  Many indirect and direct detection experiments
aim to discover and measure the properties of WIMPs~\cite{hooper}.
Direct WIMP detection experiments search for the interaction of WIMPs
with a nucleus in the detector, resulting in low-energy nuclear
recoils~\cite{Gaitskell:2004gd}.  Most experiments seek to detect the
kinetic energy deposited by the recoiling nucleus; a handful of recent
efforts, including this work, also seek to detect the direction of the
nuclear recoil, and in this way, infer the direction of incoming
WIMPs~\cite{Ahlen:2009ev,Buckland:1994gc,Alner:2005xp,newage,Dujmic:2007bd,Santos:2007ga,Naka:2007zz}.
The arrival direction of WIMPs is predicted to peak in the direction
opposite to the earth's motion around the galactic center in the
simplest dark matter halo model, and have a time-varying asymmetry
because the Earth's rotation gives angular modulation in
time~\cite{Spergel:1987kx}.  The angular signature of directional
detection offers the potential for unambiguous observation of dark
matter~\cite{Green:2006cb}.  This paper presents the first dark matter
limit from the DMTPC directional detection experiment, from a surface
run at MIT.

\section{The Dark Matter Time Projection Chamber Experiment}
DMTPC is a dark matter detector designed to measure
the direction and energy of recoiling fluorine nuclei.  CF$_4$ is
chosen as a target due to good scintillation
characteristics~\cite{Kaboth:2008mi} and the relatively large
predicted axial-vector coupling for fluorine, allowing sensitivity to
spin-dependent WIMP interactions~\cite{Jungman:1995df,lewin1995}.

\begin{figure}
\centering
\includegraphics[width=8cm]{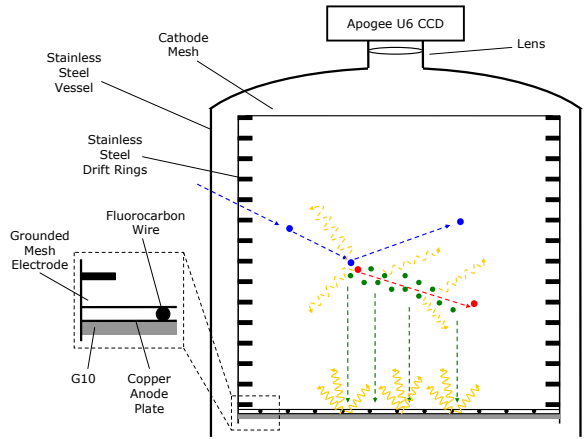}
\setlength{\abovecaptionskip}{0pt}
\caption{A schematic of one TPC (not to scale).  A WIMP (blue) induces a nuclear recoil (red) which produces ionization electrons (green) and scintillation light (yellow).  \label{fig:apparatus}}
\setlength{\belowcaptionskip}{0pt}
\end{figure}

The detector consists of two optically isolated back-to-back
low-pressure time projection chambers, with a
14.6$\times$14.6$\times$19.7~cm$^{3}$
(15.9$\times$15.9$\times$19.7~cm$^{3}$) fiducial volume for the top
(bottom) TPC.  The TPCs are filled with 75$\pm$0.1 Torr of CF$_4$
corresponding to a 3.3~g (2.85~g) fiducial mass of CF$_4$ (F).  In
75~Torr of CF$_4$, a recoiling fluorine nucleus with 50~keV kinetic
energy would travel approximately 1~mm before stopping.  The cathode
and ground planes of the TPC are 27 cm diameter meshes with 256~$\mu$m
pitch.  The grounded mesh sits 0.5~mm from the copper-clad G10 anode
plate (see Figure~\ref{fig:apparatus}).  Ionization electrons from
interactions in the fiducial volume drift in a uniform electric field
of 0.25~kV/cm towards the amplification region (14.4~kV/cm) where
avalanche multiplication amplifies the electron signal and produces
scintillation.  The wavelength spectrum of the scintillation light
peaks at $\sim$600 nm, with roughly two-thirds of the scintillation
emission in the visible~\cite{Kaboth:2008mi}.  The gas gain is
approximately $4 \times 10^4$, measured with an $^{55}$Fe calibration
source.  The operating anode voltage is chosen to maximize the gain
while limiting the rate of electronic discharge between the anode and
ground plane to $<$0.025 Hz.  The drift electric field is chosen to
minimize the transverse diffusion of the drifting electrons.  For a
more detailed discussion of the 10 liter detector amplification and
diffusion, see~\cite{Dujmic:2008ut} and~\cite{Caldwell:2009si}.

Scintillation light produced in the amplification region is focussed
by a Nikon photographic lens (f/1.2, 55~mm focal length) onto an
Apogee Alta U6 camera containing a 1024$\times$1024 element Kodak
1001E CCD chip with 24$\times$24 $\mu$m$^2$ pixels.  The CCD clock
rate is 1 MHz with 16-bit digitization, and typical readout time is
0.2 seconds.  With this camera we are read-noise limited.  To improve
the signal-to-noise ratio (and to reduce deadtime from CCD readout),
pixels are binned 4$\times$4 prior to digitization.  In addition to
optical readout, we also digitize the integrated charge induced on the
anode, although we do not use the charge data in this analysis.

The surface run data set consists of 231,000 five-second CCD exposures
from each camera, collected without trigger or camera shutter.  Of
these, 10.5\% and 4.4\% are rejected by analysis cuts as spark events
in the top and bottom TPCs respectively, and 3.5$\times$10$^{-3}$\%
and 8.7$\times$10$^{-4}$\% as the associated residual bulk image (RBI)
background pixels (described in Section~\ref{subsec:bgnds})
respectively.  After correcting the live time for these analysis cuts,
the data set live time averaged over the two cameras is 12.35 days.
This does not include parasitic exposure (when pixels are exposed
during CCD readout and event writing), which increases the live time
by approximately 11\%.  The data taking efficiency was approximately
65\%, including time for gas refilling.

\subsection{Calibration and Reconstruction}
Track length and energy calibrations employ $^{241}$Am alpha sources
at fixed locations in the top and bottom TPCs.  The calibration
sources are placed inside each TPC, on the field cage rings.  The
energies of the two sources, used for top and bottom TPC calibrations,
are 4.51$\pm$0.05 MeV and 4.44$\pm$0.05 MeV respectively.  These
energies come from independent measurements of each source with a
Canberra 450-20AM surface barrier detector, calibrated with decay
alphas from radon-enriched $N_2$ gas.  The energies are slightly
different for the two sources, likely because of differences in
thickness of the thin gold windows of each of the source holders.  The
length calibration relies on the known horizontal separation,
2.5$\pm$0.1 cm, of resistive separators in the amplification region,
shown in Figure~\ref{fig:apparatus}.  Tracks from alphas have
decreased light yield at the spacer locations.  Fitting Gaussian
profiles to these regions gives the spacer positions in the image
plane, and shows that each 24$\mu$m$\times$24$\mu$m CCD pixel images
143$\pm$3~$\mu$m$\times$143$\pm$3~$\mu$m
(156$\pm$3~$\mu$m$\times$156$\pm$3~$\mu$m) of the top (bottom) anode.
\begin{figure} 
\centering
\includegraphics[height=7cm, angle=90]{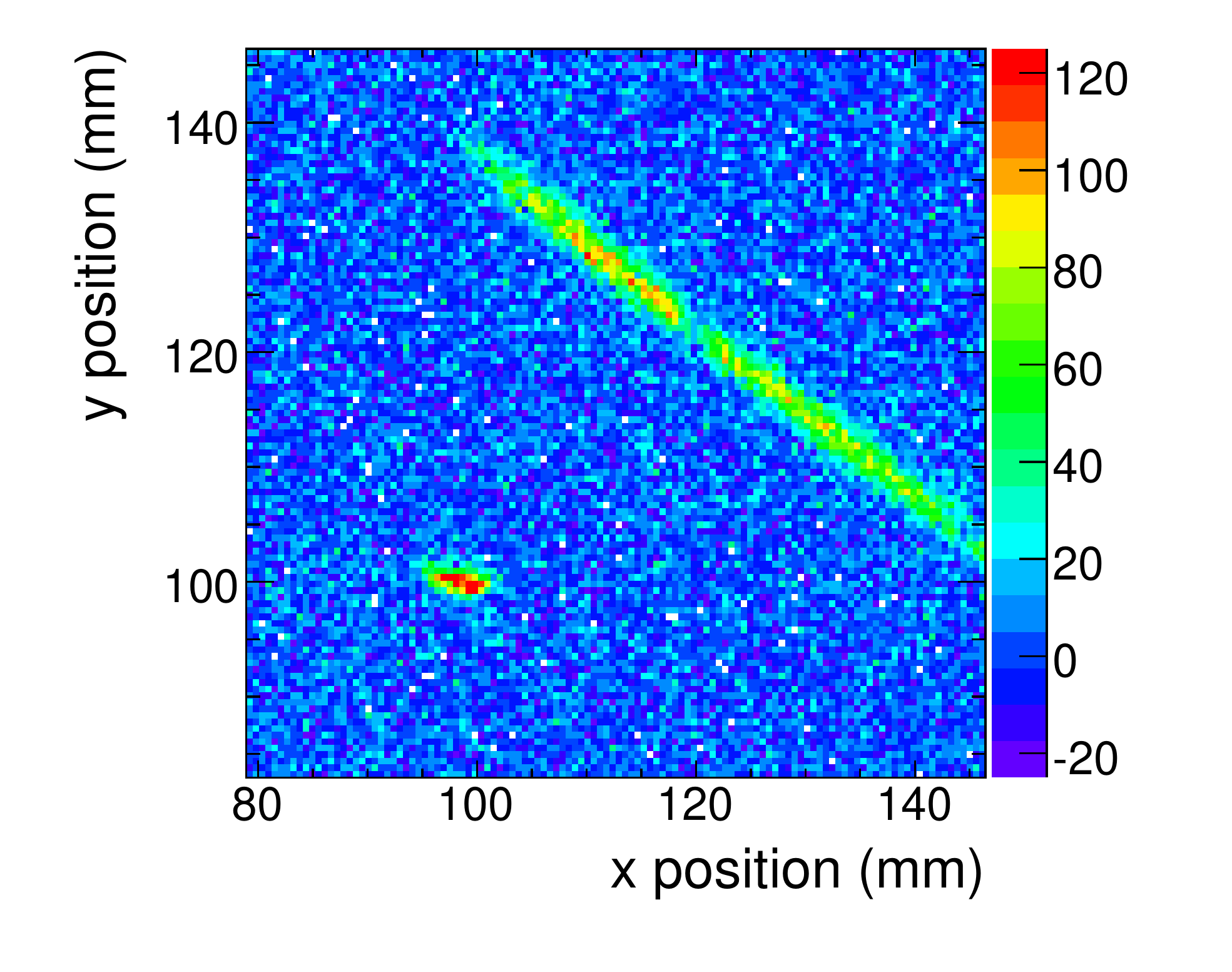}
\setlength{\abovecaptionskip}{0pt} 
\caption{A sub-region of a background-subtracted event from the top camera containing a segment of an alpha track (long) and a candidate nuclear recoil (short); intensity is in units of ADU, indicated by color (white pixels have $<$-20 ADU).  Both tracks exhibit an asymmetry of energy loss along the axis of the track, consistent with the Bragg curve. \label{fig:event}}
\setlength{\belowcaptionskip}{0pt} 
\end{figure}

The energy response of the detector is obtained from the same data.
The integral light yield of segments of alpha tracks at known
distances from the source, in the arbitrary digital units (ADU) of the
CCD, is compared to the SRIM simulation~\cite{srim} prediction for the
visible energy loss in that segment.  The segment length is chosen such
that the SRIM prediction for the energy loss in each segment is
100-1000 keV, depending on the location and size of the segment along
the alpha track.  This procedure is done in the region of the alpha
track where the alpha energy is above 1 MeV (before the Bragg peak).
According to SRIM, at these energies, the alpha energy loss is $>$97\%
electronic and so we are not sensitive to assumptions about the
nuclear quenching in this calibration.  This procedure gives the
energy calibration of 9.5$\pm$0.5 and 12.9$\pm$0.7 ADU/keV
respectively for the top and bottom cameras.  After accounting for the
different conversion gain and read noise for each camera, the
signal-to-noise is approximately the same between the two.  The
uncertainty on the calibration is estimated by varying the size and
location of the alpha segments relative to the start of the alpha
track.
\begin{figure*}
\centering
\includegraphics[height=6.5cm, angle=90]{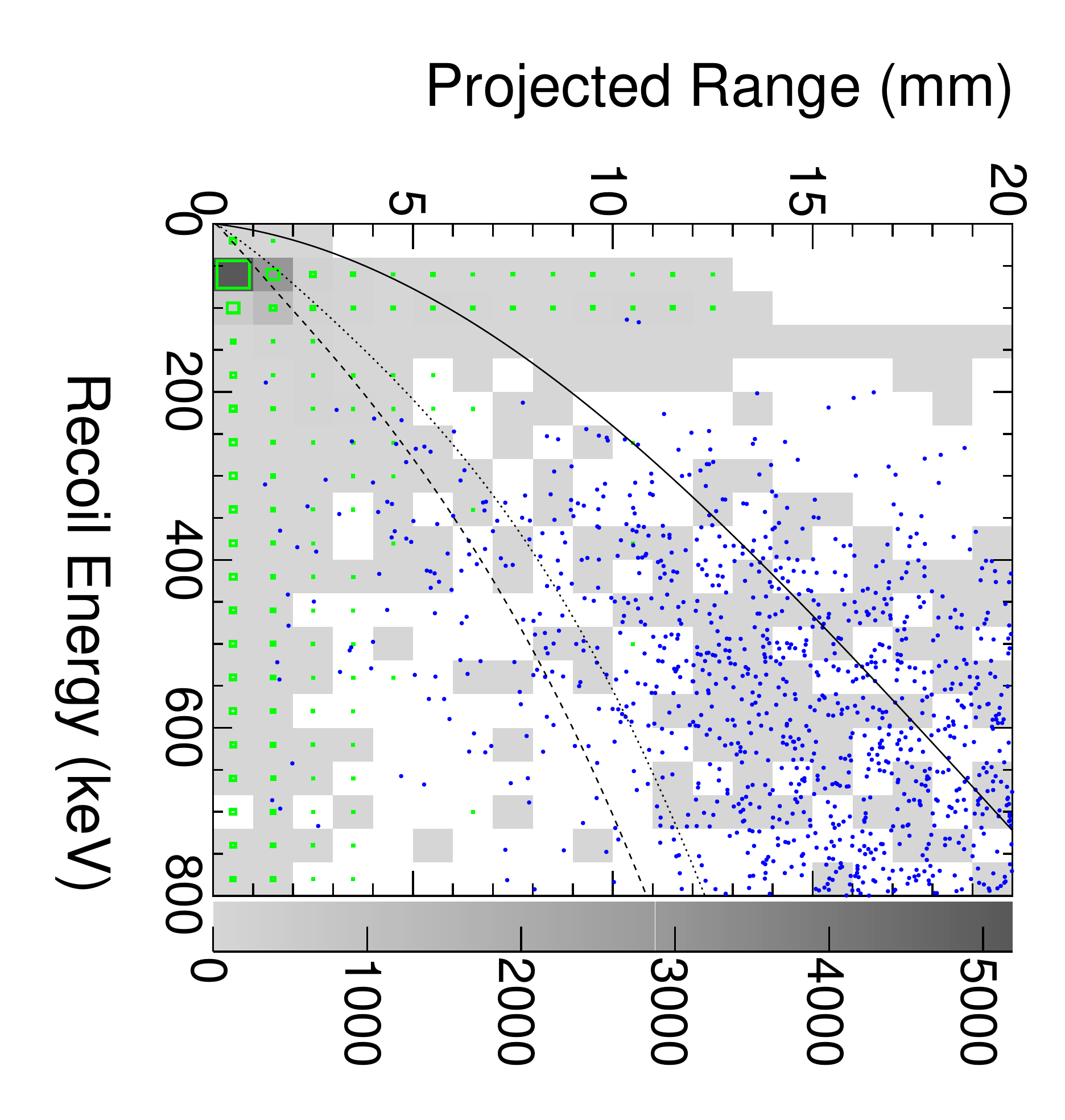}
\includegraphics[height=6.5cm, angle=90]{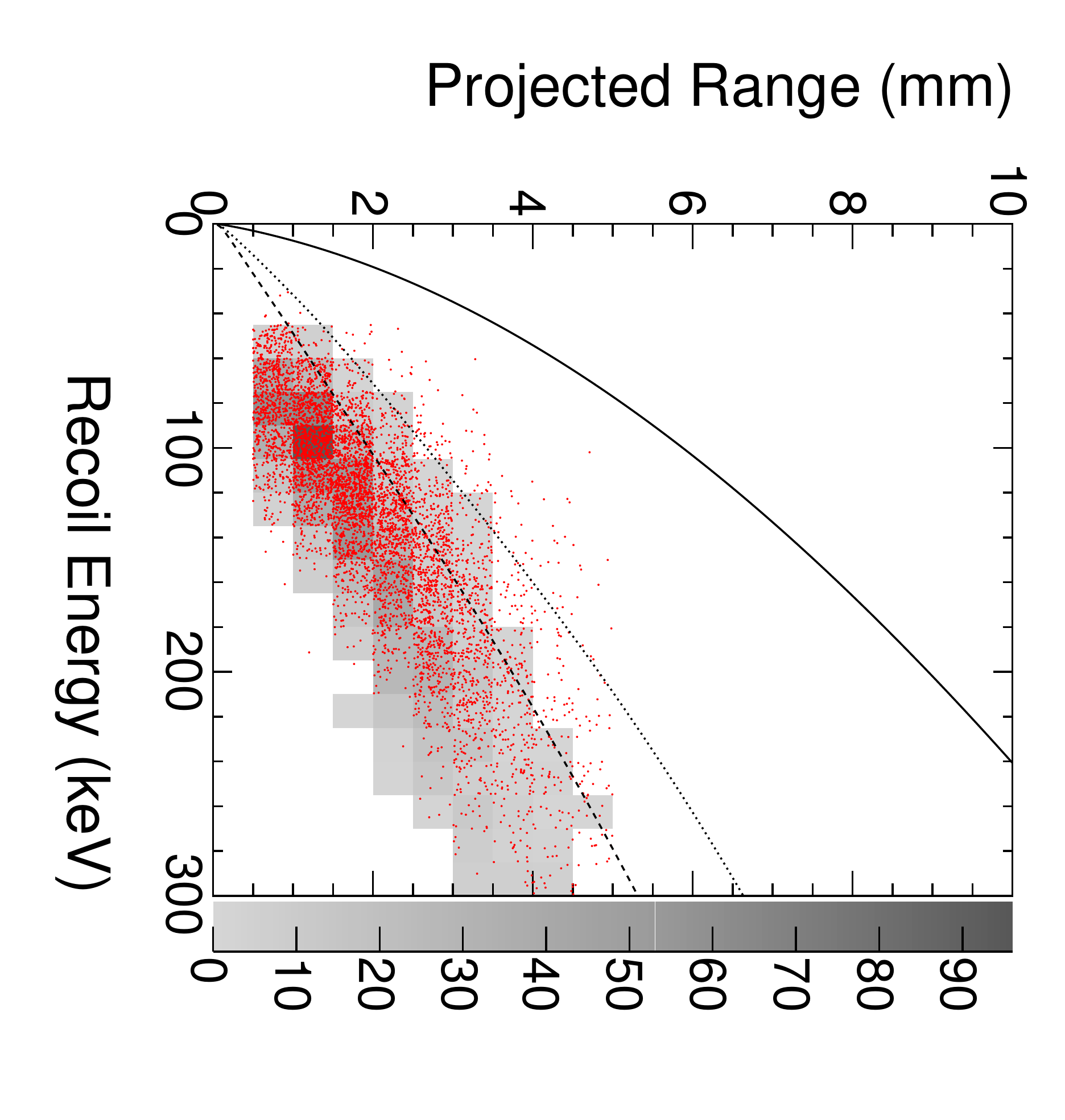}
\setlength{\abovecaptionskip}{0pt}
\caption{Reconstructed projected range (mm) vs. reconstructed energy (keV).  Left: the background populations; RBIs (shaded squares), alphas (blue points), and CCD interactions (green open boxes).  Right: $^{252}$Cf calibration data (red points) compared with $^{252}$Cf-F Monte Carlo (shaded squares) after nuclear recoil selection cuts.  Lines are SRIM predictions for the maximum projected range vs. recoil energy for helium (solid), carbon (dotted), and fluorine (dashed) nuclei.  \label{fig:rve_1}}
\setlength{\belowcaptionskip}{0pt}
\end{figure*}
\begin{figure*}
\centering
\includegraphics[height=6.5cm, angle=90]{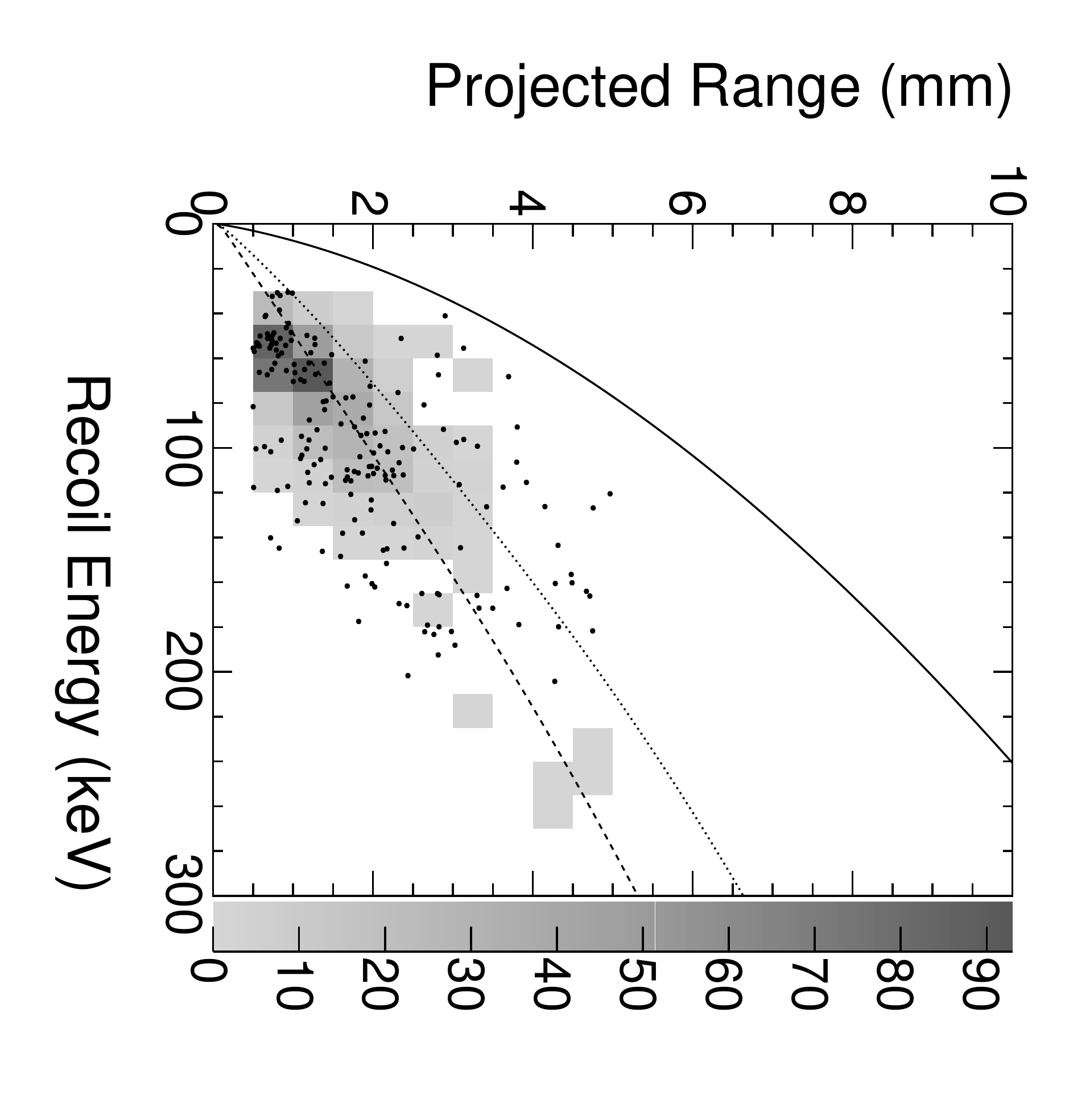}
\includegraphics[height=6.5cm, width=6.25cm, angle=90]{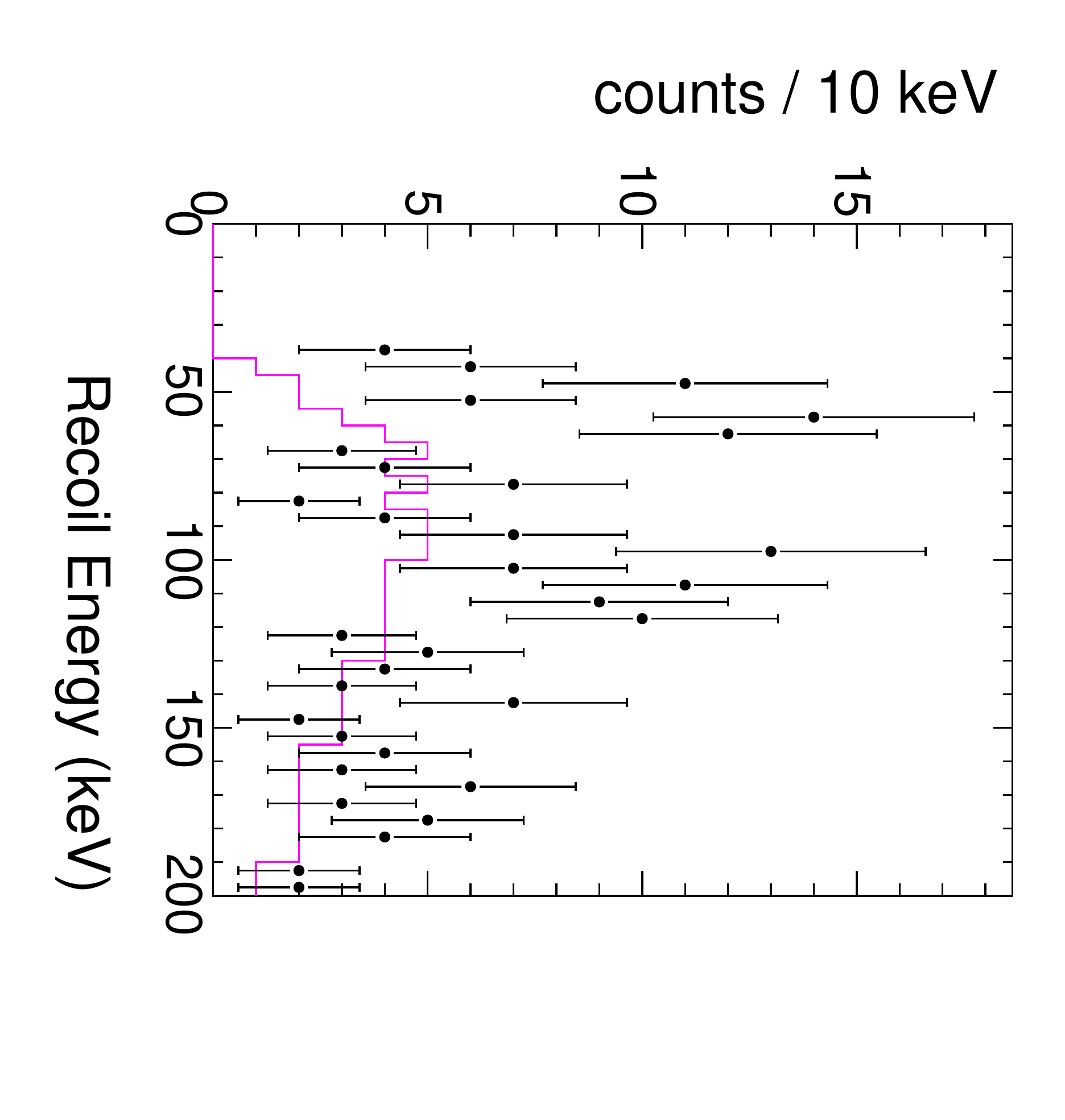}
\setlength{\abovecaptionskip}{0pt}
\caption{Left: reconstructed projected range (mm) vs. reconstructed energy (keV) for WIMP-search data (black points) compared with 200 GeV/c$^2$ WIMP-F Monte Carlo (shaded squares) after nuclear recoil selection cuts.  Lines are SRIM predictions for the maximum projected range vs. recoil energy for helium (solid), carbon (dotted), and fluorine (dashed) nuclei.  Right: reconstructed energy (keV) for WIMP-search data passing nuclear recoil selection cuts (black points with statistical errors), and the predicted neutron-induced background (magenta line).    \label{fig:rve_2}}
\setlength{\belowcaptionskip}{0pt}
\end{figure*}

The system gain (ADU/keV) may vary with position and in time.  The
gain non-uniformity across the field of view due to local variations
of the amplification is measured with a 14 $\mu$Ci $^{57}$Co source,
which provides uniform illumination of the field-of-view from
scattered 122 keV photons.  The track-finding algorithm does not
identify distinct tracks in the $^{57}$Co data, largely because these
fail the requirement of having at least five contiguous pixels above
threshold (described further below).  This is consistent with the
predicted low ionization density of electron-like tracks (see, for
example Figure 16 in \cite{Ahlen:2009ev}).  Rather, these events may
have a few pixels above background in the entire field of view.  To
obtain high statistics, each gain non-uniformity measurement is
integrated over 10,000 seconds; from calculation, the intensity and
position of the source are such that the area imaged by each CCD pixel
is covered by at least one electron recoil per second.  The measurement yields
a 10$\%$ variation of the total system gain, which is included as a
position-dependent correction in the gain systematic study in
Section~\ref{subsec:bgnds}.  The stability of the gain vs. time was
measured to be 1$\%$ over 24 hours using an $^{241}$Am alpha source.
To maintain 1\% gain uniformity, the chamber is evacuated to 10 mTorr
and refilled with CF$_4$ every 24 hours.

In the track reconstruction, raw CCD images are first background
subtracted using an average of 100 dark frames (taken with the shutter
closed) to remove spatial non-uniformities in the CCD dark rate.  Dark
frames are collected every 1000 exposures; $\ge$10$\sigma$ lone
outlier pixels are replaced with the mean of the 8 neighboring pixels
before averaging.  After dark frame subtraction, the same lone pixel
cleaning procedure is applied to each image.  For events which are not
classified as sparks, any residual mean pixel count is removed by
subtracting the mean pixel intensity of the image.
The track finding algorithm bins the CCD images in software to
8$\times$8 pixels.  Groups of five contiguous bins with at least
$3.7\sigma$ counts per bin are identified as clusters; 3.7 is chosen
to optimize the energy reconstruction resolution.  Clusters which lie within
three bins of each other are combined, to account for the resistive
separators segmenting tracks.  Example tracks are shown in Figure
\ref{fig:event}.  Reconstructed quantities are determined from the
original pixels (4$\times$4 binned in readout) in each cluster.

The visible track energy is determined by the integral of the counts
in a track, divided by the energy calibration constant (ADU/keV).  To
convert visible energy to nuclear recoil energy (shown in Figures
\ref{fig:rve_1}-\ref{fig:data_er_phired}) we use the CF$_4$
quenching factor calculated in \cite{Hitachi:2008kf}, and the SRIM
prediction of nuclear and electronic energy loss for fluorine.  The
projected range of a track on the image plane is calculated as the
distance between the maximally separated pixels in the track with
yield $>$3.7 $\sigma$ above the image mean, multiplied by the length
calibration constant ($\mu$m/pixel).  The track angle in the
amplification plane ($\phi$) is determined by finding the major axis angle of
an ellipse with the same second moment as the pixels in the cluster.
The sense of the direction is estimated from the skewness of the track
light yield.  

The recoil energy and angle reconstruction resolution are 15\% and
40$^o$ at 50 keV visible energy (80 keV nuclear recoil energy).  The
energy resolution is measured with alpha calibration data (shown in
\cite{Caldwell:2009si}, Figure 2), and the energy resolution in Monte
Carlo is validated by comparison with the measured energy resolution
in alpha track segments.  From nuclear recoil Monte Carlo, the energy
resolution varies with energy as $(\sigma_E/E_{vis})^2 \ = \ a^2 +
(b/\sqrt{E_{vis}})^2 + (c/E_{vis})^2$ where $a$ = 0.051, $b$ =
3.8$\times$10$^{-3}$, and $c$ = 6.1, for events passing the nuclear
recoil selection cuts described in the following section.  The angular
resolution is estimated with a Monte Carlo simulation of fluorine
recoils from the $^{252}$Cf calibration source.  The simulation is
based on measured detector
characteristics~\cite{Kaboth:2008mi,Caldwell:2009si}; the $^{252}$Cf-F
Monte Carlo is compared with data in Figure~\ref{fig:rve_1}.  From
Monte Carlo, the direction reconstruction resolution varies with
energy as $(\sigma_{\phi}/E_{vis}) \ = \ a \ exp(-E_{vis}/b)$ where
$a$ = 6.1 and $b$ = 24.3, for events passing the nuclear recoil
selection cuts described in the following section.  More detail on
directionality studies with this detector technology can be found
in~\cite{Dujmic:2008ut}.

\subsection{Surface Run Results \label{subsec:bgnds}}
A major goal of the surface run was to identify detector backgrounds
prior to underground operations.  We found two broad categories:
events which produce ionization outside the TPC drift volume, and
events which occur inside it.  A summary is given in Table
\ref{table:rates}.

Background events producing ionization outside the fiducial volume are
mostly interactions of cosmic rays or radioactivity in the CCD chip,
which is a well documented phenomenon~\cite{ccds}.  These may be
removed in the future by requiring coincidence of CCD and charge or PMT
readout; in this CCD-only analysis, we reject these events in
software.  Such tracks typically have a few bins with very high
yields.  We identify these events by the large ADU and RMS of the
pixels comprising the track
.  Another type of outside event is associated with sparks in the
amplification region.  Sparks are identified by having an image mean
which differs by $>$1\% from the previous image.  For comparson,
images containing very bright alpha tracks differ in this metric by
$<$0.01\%.  Sparks may induce residual bulk images (RBIs), which
appear at the same spatial position for many subsequent images.  RBIs
are the result of the leakage of charge from the epitaxial/substrate
interface of the CCD; these are a well-known background in
front-illuminated CCDs associated with interactions of $>$600 nm
photons in the chip~\cite{ccds,rbis}.  We identify these events by
their coincident positions.
\begin{table}
\centering
\begin{tabular}{cc}
\hline
\hline
Event Selection Cut & Rate (Hz) \\
\hline
All Tracks & 0.43\\
Residual Bulk Images & 0.15\\
CCD Interactions & 4.4$\times 10^{-3}$\\
Alpha Candidates & 8.2$\times 10^{-5}$\\
Nuclear Recoil Candidates in 80 $< E_R <$ 200 keV & 5.0$\times 10^{-5}$ \\
\hline
\hline
\end{tabular}
\caption{Surface run event rates (Hz) after each background rejection cut, summed over the two cameras. 
\label{table:rates}}
\end{table}

Background events producing ionization inside the fiducial volume come
primarily from alphas and neutrons.  Alpha particles are emitted by
radio-impurities in or on the materials of the detector; the majority
are from the stainless steel drift cage.  These are identified as CCD
edge-crossing tracks.  Another characteristic of alphas is their long
range; we require nuclear recoils to have projected ranges $<$5 mm.
This range vs. energy discrimination is unique to tracking detectors.
Figure~\ref{fig:rve_1} (left) shows events identified as alpha
particles in comparison with the SRIM prediction for the maximum
projected range vs. visible energy; tracks which are not parallel to
the image plane have projected ranges less than this maximum.  The
ambient neutron flux comes from $^{238}$U and $^{232}$Th decays, and
from cosmic ray spallation. Figure~\ref{fig:rve_1} (right) shows
calibration $^{252}$Cf neutron-induced recoils, which are
indistinguishable from a dark matter signal on an event-by-event basis
in range vs. energy; these tracks sample a range of angles relative to
the amplification plane and are shown compared with the SRIM
prediction for the maximum projected range vs. energy and to the
detector simulation.  There is no evidence for gamma-induced electron
backgrounds~\cite{Dujmic:2008iq}; the measured rejection is
$>10^{6}$~\footnote{The study in~\cite{Dujmic:2008iq} used an 8
$\mu$Ci $^{137}$Cs calibration source, positioned on the drift mesh
inside the detector, for 1/3 day.  No tracks were identified by the
reconstruction algorithm; the rejection factor comes from comparing
the observed with the expected number of 661 keV betas associated with
the $^{137}$Cs decay.  From simulation, we find that betas do not
satisfy the track-finding algorithm requirement of 5 contiguous pixels
with signal-to-noise ratio above the 3.7$\sigma$
threshold~\cite{Ahlen:2009ev}.}.  The events remaining after all
background cuts are shown in Figure~\ref{fig:rve_2} (left), compared
to WIMP Monte Carlo.
\begin{figure*}
\includegraphics[height=6.5cm, width=6.5cm, angle=90]{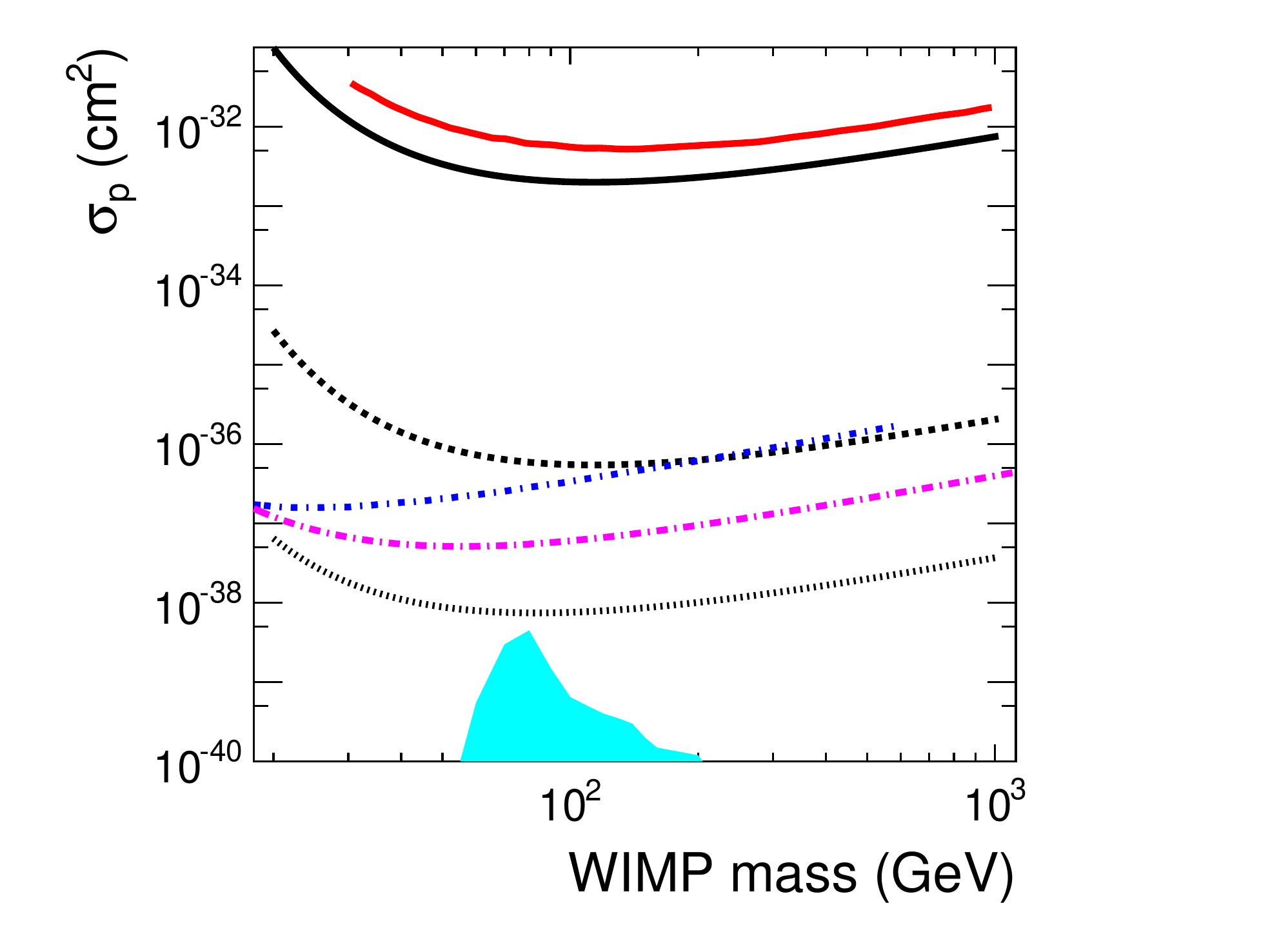}
\includegraphics[height=6.5cm, width=6.5cm, angle=90]{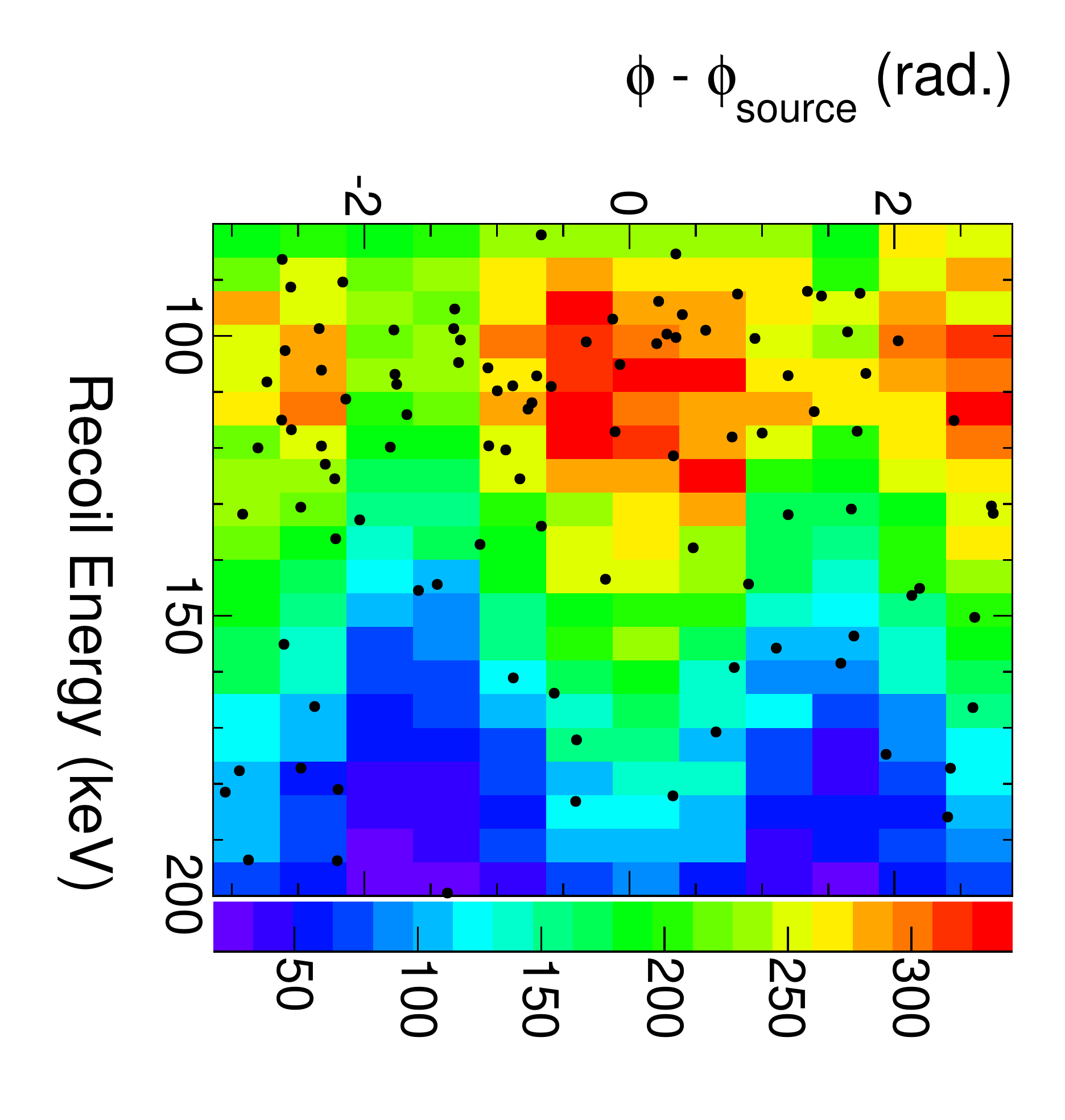}
\setlength{\abovecaptionskip}{0pt}
\caption{Left: 90\% confidence level limit on the spin-dependent WIMP-proton cross section vs. dark matter particle mass from DMTPC surface data (black solid line), compared with the NEWAGE~\cite{newage} underground directional result (red solid line), and the two leading limits from conventional detectors, COUPP~\cite{coupp} (magenta dash-dotted line) and PICASSO~\cite{picasso} (blue dash-dotted line).  The cyan shaded region shows MSSM parameter space~\cite{Ellis:2000jd}.  The projected sensitivity for DMTPC at WIPP, with 1 year exposure (black dashed line), and a 1 $\rm m^3$ detector at WIPP with 50 keV energy threshold (black dotted line), are also shown. Right: reconstructed angle relative to source $\phi-\phi_{source}$ (radians) vs. recoil energy (keV) for data passing nuclear recoil selection cuts in the dark matter search energy range; WIMP search data (black points) is compared with $^{252}$Cf data (color indicates number of events).  For WIMP search data $\phi_{source}$ is the direction to Cygnus, for $^{252}$Cf data $\phi_{source}$ is the direction to the source in the laboratory (which is effectively a point source).  \label{fig:data_er_phired} }
\setlength{\belowcaptionskip}{0pt}
\end{figure*}

We set a limit on the spin-dependent WIMP-proton interaction cross
section using the method described in \cite{lewin1995}.  The signal
efficiency is calculated from the WIMP Monte Carlo simulation.  The
analysis energy window, 80-200 keV, is chosen to maximize the integral
above threshold of the product of efficiency and predicted
WIMP-induced recoil spectrum (for $m_{WIMP}$ = 200 GeV/c$^2$); this
averaged efficiency is maximized at 70\% at 80 keV threshold energy.
There are 105 events after all cuts in 80 $<$ $E_{recoil}$ $<$ 200
keV, with 74 predicted neutron background events in this window based
on the surface neutron spectrum measurement in~\cite{neutrons}
(Figure~\ref{fig:rve_2}, right).  We do not take into account the
building around the detector, and so assign 100\% uncertainty to the
neutron background and report the limit assuming zero expected events.
Using the Feldman-Cousins method~\cite{Feldman:1997qc}, we set a 90\%
confidence level limit on the spin-dependent WIMP-proton cross
section, shown in Figure~\ref{fig:data_er_phired} (left).
Following~\cite{lewin1995}, we use the thin-shell spin-dependent form
factor approximation, and the interaction factor $C_{Wp}^2$ = 0.46 for
Higgsino-proton coupling.  The 90\% C.L. cross section upper limit is
$2.0\times10^{-33}$ cm$^{2}$ at 115 GeV/c$^2$ WIMP mass.  If we vary
the gain non-uniformity by 100\%, the limit is $<2.3\times10^{-33}$
cm$^{2}$.  If we include the estimated background of 74 events, the
limit is $<8.0\times 10^{-34}$ cm$^2$.

We evaluate the probability that events passing the nuclear recoil
selection cuts come from an isotropic background vs. anisotropic
WIMP-induced recoil angle distribution.  The Rayleigh statistic is a
powerful tool to analyze the uniformity of a distribution of angles
when looking for a preferred direction~\cite{mardia2000directional}.
Using the Rayleigh statistic, we quantify the anisotropy in
$(\phi-\phi_{source})$, which is the most sensitive variable to test for
anisotropy in the case of two dimensional
readout~\cite{morgan2005directional}.  $(\phi-\phi_{source})$ is the
difference between the reconstructed $\phi$ and the projection of the
expected dark matter direction at the time of each event onto the
image plane.  The $(\phi-\phi_{source})$ vs. $E_R$ distribution after
nuclear recoil selection cuts in is shown in Figure
\ref{fig:data_er_phired} (right).  We find no statistically
significant deviation from a uniform distribution; 36\% of the time
uniformly distributed data have a Rayleigh value higher than that of
our candidate events.  The reconstructed angle of the $^{252}$Cf
calibration data relative to its source is also shown in Figure
\ref{fig:data_er_phired} (right).  The $^{252}$Cf calibration source is effectively a point source in the lab frame at $\phi_{source}$=0.  The Rayleigh test applied to the $^{252}$Cf calibration data after the nuclear recoil selection
cuts, in the same recoil energy range (80-200 keV), gives a probability of
$<$1\% for a uniform distribution.

\section{Conclusions}  
We present the first dark matter limit from DMTPC, $\sigma_{\chi-p}$
$<2.0\times10^{-33}$ cm$^{2}$ at 90\% C.L., from a 35.7 g-day surface
exposure of a 10 liter detector.  The 10$^4$ rejection of backgrounds
using range vs. energy properties of nuclear recoils, from Table
\ref{table:rates}, is an impressive demonstration of the low pressure
directional time projection chamber concept.  We find that the
backgrounds in the analysis window of 80-200 keV are qualitatively
consistent with the predicted neutron background.  The 10L detector
described here began running underground at the Waste Isolation Pilot
Plant outside Carlsbad, NM in October 2010.  The depth of the WIPP
site is 1.6 km water-equivalent.  The gamma, muon, and radon
background levels have been measured, and the neutron background has
been estimated at this site~\cite{Esch:thesis}.  Based on these, we
project that underground operation will lower the expected neutron
background to $<$1 event/year.  The projected zero background
sensitivity of this detector at WIPP for a 1 year exposure is shown in
Figure~\ref{fig:data_er_phired} (left).  DMTPC has built a
second-generation detector with radio-pure materials for operation at
WIPP; this is expected to substantially reduce alpha backgrounds, and
fiducial volume coverage by CCDs in coincidence with charge readout
will eliminate CCD backgrounds.  At the scale of a 1 $\rm m^3$
detector (300 g target), which the collaboration is actively
developing, this detector technology is competitive with the best
current spin-dependent cross section limits from conventional dark
matter detectors, also shown in Figure \ref{fig:data_er_phired}
(left).

\begin{acknowledgments}
We acknowledge support by the Advanced Detector Research Program of
the U.S. Department of Energy (contract number 6916448), as well as
the Reed Award Program, the Ferry Fund, the Pappalardo Fellowship
program, the MIT Kavli Institute for Astrophysics and Space Research,
the MIT Bates Research and Engineering Center, and the Physics
Department at the Massachusetts Institute of Technology.  We would
like to thank Mike Grossman for valuable technical assistance.

\end{acknowledgments}

\bibliography{note031}

\end{document}